\documentclass{llncs}
\usepackage{amsmath,amssymb,amsfonts}
\usepackage{hyperref}
\usepackage{algorithmic}
\usepackage[numeric]{amsrefs}
\usepackage{rotating} 
\input colordvi
\usepackage{graphicx}

\newcommand{\secref}[1]{Section~\ref{#1}}

\numberwithin{equation}{section}

\newcommand\g{\gamma}

\renewcommand\L{\Lambda}

\newcommand\G{\Gamma}
\newcommand\f{\frac}

\newcommand{\Z}{{\mathbb{Z}}}
\newcommand{\R}{{\mathbb{R}}}

\renewcommand\({\left(}
\renewcommand\){\right)}
\newcommand{\ttwo}[4]{
\(\begin{smallmatrix}{#1} & {#2}
\\ {#3} & {#4} \end{smallmatrix}\)}

\newcommand{\gobble}[1]{}
  \newcommand{\rangeref}[2]{%
    \ref{#1}--\afterassignment\gobble\fam 0\ref{#2}%
  }

\makeatletter
\def\imod#1{\allowbreak\mkern5mu({\operator@font mod}\,#1)}
\makeatother

\begin{document}

\title{Generating cryptographically-strong random lattice bases and recognizing rotations of $\mathbb{Z}^n$}

\author{Tamar Lichter Blanks\thanks{Supported by a National Science Foundation Graduate Research Fellowship.}\and Stephen D.~Miller\thanks{Supported by National Science Foundation Grants CNS-1526333 and CNS-1815562.}
}

\institute{Department of Mathematics, Rutgers University}

\date{February 11, 2021}

\maketitle

\begin{abstract}
Lattice-based cryptography relies on generating random bases which are difficult to fully reduce.  Given a lattice basis (such as the private basis for a cryptosystem), all other bases are related by multiplication by matrices in $GL(n,\mathbb{Z})$.  We compare the strengths of various methods to sample random elements of $GL(n,\mathbb{Z})$, finding some are stronger than others with respect to the problem of recognizing rotations of the $\mathbb{Z}^n$ lattice.  In particular, the standard algorithm of multiplying unipotent generators together (as implemented in Magma's {\tt RandomSLnZ} command) generates instances of this last problem which can be efficiently broken, even in dimensions nearing 1,500.  Likewise, we find that the random basis generation method in one of the NIST Post-Quantum Cryptography competition submissions (DRS) generates instances which can be efficiently broken, even at its 256-bit security settings.  Other random basis generation algorithms (some older, some newer) are described which appear to be much stronger.

\vspace{.2cm}

\end{abstract}

\section{Introduction}\label{sec:intro}

In  cryptography one often encounters problems which are easy to solve using a secret private basis of a lattice $\L\subset \R^n$, but are expected to be difficult to solve using suitably-chosen public bases.  Famous examples include the Shortest Vector Problem (SVP) and Closest Vector Problem (CVP).

 In \cite{LenstraSilverberg} Lenstra and Silverberg posed the challenge of whether highly-symmetric lattices  have hard bases, and proved several interesting results along these lines (related to earlier work of Gentry-Szydlo \cite{GentrySz}; see also \cite{LenstraSilverberg0,LenstraSilverberg2}).  One particularly beautiful question they posed is:
\begin{equation}\label{LSQ}
  \text{can one efficiently recognize rotations of the standard $\Z^n$ lattice?}
\end{equation}
To be more precise, this problem can be stated in two different group-theoretic ways (the second being the formulation in \cite[\S 2]{LenstraSilverberg}).  Let $\{b_1,\ldots,b_n\}$ denote a basis for $\L$ and let $B$ denote the $n\times n$ matrix whose $i$-th row is $b_i$:
\begin{center}
\text{
\fbox{\fbox{\begin{minipage}{11.5cm}
\begin{algorithmic}
\STATE {\bf \underline{Problem 1a (Decision version).}}
\vspace{.15cm}
\STATE Can one efficiently factor $B$ as $B=MR$, with $M\in GL(n,\Z)$ and $R\in O(n)$?
\STATE
\STATE {\bf \underline{Problem 1b (Search version).}}
\vspace{.15cm}
\STATE If so, efficiently find such matrices $M\in GL(n,\Z)$ and $R\in O(n)$.
\end{algorithmic}
\end{minipage}}}}
\end{center}

\noindent  Alternatively, following \cite{GeisslerSmart} and \cite[\S 2]{LenstraSilverberg} we may suppose one is given a  positive-definite symmetric matrix $G\in SL(n,\Z)$ (which we think of as the Gram matrix $G=BB^t$ of $\L$):
\begin{equation}\label{LSQGabnew}
\text{
\fbox{\fbox{\begin{minipage}{11.5cm}
\begin{algorithmic}
\STATE {\bf \underline{Problem 2a (Decision version).}}
\vspace{.25cm}
\STATE Given a positive-definite integral matrix $G$, efficiently determine whether or not there is some $M\in GL(n,\Z)$ such that $G=MM^t$.
\STATE
\STATE {\bf \underline{Problem 2b (Search version).}}
\vspace{.25cm}
\STATE If so, efficiently find such a matrix $M\in GL(n,\Z)$.
\end{algorithmic}
\end{minipage}}}}
%
\end{equation}
Clearly, Problem 1 reduces to Problem 2 with $G=BB^t$.  Conversely, one can orthogonally diagonalize the matrix $G$ in Problem 2 as $G=PDP^t$ for some $P\in O(n)$ and diagonal matrix $D$ with positive diagonal entries.  Then $B=PD^{1/2}$ solves the equation $G=BB^t$, and Problem 2 therefore reduces to Problem 1 (modulo  technicalities we will not delve into, such as that the entries of $P$, $D$, and $B$ may in general be irrational).

In particular, by orthogonal diagonalization it is trivial to find  a non-integral solution $M\in GL(n,\R)$ to Problem 2.     However, imposing the constraint that $M\in GL(n,\Z)$ adds an intricate dose of number theory, since Problem 2a then becomes a class number problem:~indeed, in large dimensions $n$ there is a combinatorial explosion of possible $GL(n,\Z)$-equivalence classes.\footnote{For example, the $E_8$ lattice has a Gram matrix $G$ in $SL(8,\Z)$, but is not isometric to the $\Z^8$ lattice.  In general the number of $GL(n,\Z)$-equivalence classes of such integral unimodular lattices grows  faster than exponentially in $n$ \cite[Chapter 16]{CS}.}

Both Problems 1 and 2 have inefficient solutions using sufficiently strong lattice basis reduction.  For example, the given information is sufficient to determine whether or not all lattice vector norms are square-roots of integers, and an SVP solver can determine  the shortest nonzero norm $\lambda_1(\L)$.  If $\lambda_1(\L)\neq 1$, the lattice $\L$ is definitely not a rotation of $\Z^n$ and Problems 1a and 2a have negative solutions.  However, if one finds a vector of norm 1 and all lattice norms are square-roots of integers, it is then easy to see (by subtracting multiples of this vector to obtain an orthogonal complement) that the dimension in Problems 1b and 2b reduces from $n$ to $n-1$.
It was recently shown in \cite{recentlyshown} that Problem 2a is in the  class NP$\cap$co-NP, using results of Elkies \cite{elkies} on characteristic vectors of lattices (see also \cite[\S9.6]{Gerstein}).

This paper primarily concerns Problem 2b, i.e., one is handed a matrix of the form $MM^t$   and wishes to efficiently recover $M$.  Of course permuting the columns of $M$ does not change $MM^t$, nor does  multiplying any subset of columns by $-1$; thus we look for solutions up to such signed permutations of the columns.  (For this reason it is equivalent to insist that $M\in SL(n,\Z)$.)   We find that the choice of procedure to  randomly generate instances of $M$ has a drastic impact on the difficulty of the problem.  We state this in terms of a probability density function $p:GL(n,\Z)\rightarrow \R_{\ge 0}$ (i.e., $\sum_{M\in GL(n,\Z)}p(M)=1$):

\begin{center}
\text{
\fbox{\fbox{\begin{minipage}{11.5cm}
\begin{algorithmic}
\STATE {\bf \underline{Problem 3 (Average case version of Problem 2b).}}
\vspace{.25cm}
\STATE Given a random matrix $M\in GL(n,\Z)$  drawn    with respect  to the probability density $p$, efficiently recover  $M$  from $MM^t$  (up to signed  permutations of the columns) with high probability.
\end{algorithmic}
\end{minipage}}}}
%
\end{center}

In \secref{sec:randomGLNZ} we compare various  methods of generating random bases of a lattice, corresponding  to different probability densities $p$ (generalizing \cite[\S 5.1.2]{AEN}; see also Section~\ref{sec:DRS}).   Here one seeks distributions for which Problem 3 is hard on average, much like SIS and LWE are average-case hard instances of variants of SVP and CVP, respectively.  We then perform experiments on them in \secref{sec:experiments}.  Some of the methods we describe, such as the long-known Algorithm 4 (see, for example, \cite{CHKPeik}), give relatively hard instances of Problem 3.  However, our main finding   is that a certain well-known existing method, namely   generating matrices by multiplying unipotents (e.g., Magma's {\tt RandomSLnZ} command), is cryptographically weak:~we were able to recover $M$ in instances in dimensions nearly 1500 (in some measurable ways these instances are comparable to NTRU lattices having purported 256-bit quantum  cryptographic strength).  That gives an example of an average-case easy distribution. In \secref{sec:DRS} we similarly find that the random basis generation method used in the DRS NIST Post-Quantum Cryptography submission \cite{DRSc} also gives weak instances of Problem 3:~in 708 hours we could recover $M$ generated using DRS's 256-bit security settings.

{\bf Acknowledgements:} it is a pleasure to thank   Huck Bennett, Leo Ducas, Nicholas Genise, Craig Gentry,  Shai Halevi, Nadia Heninger, Jeff Hoffstein,  Hendrik Lenstra,  Amos Nevo, Phong Nguyen, Chris Peikert, Oded Regev, Ze'ev Rudnick, Alice Silverberg,  Damien Stehl\'e, Noah Stephens-Davidowitz, and Berk Sunar for very helpful discussions.  We are particularly indebted to Joe Silverman for kindly suggesting an earlier variant of Algorithm 4, which is very similar  to the one we suggest here, and to Daniel J.~Bernstein for important comments about the poor equidistribution provided by Algorithm 2.  We are also grateful to Galen Collier of the Rutgers University Office of Advanced Research Computing for his assistance, and  to the Simons Foundation for providing Rutgers University with Magma licenses.

\section{Choosing random elements of $GL(n,\Z)$}\label{sec:randomGLNZ}

We consider the problem of uniformly sampling matrices in a large box\footnote{One can consider other shapes, such as balls; boxes are convenient for our applications and for making more concise statements.  The same problem for $SL(n,\Z)$ is of course equivalent.}
\begin{equation}\label{GammaT}
  \Gamma_T \ \ := \ \ \left\{M=(m_{ij}) \in GL(n,\Z) \ : \ |m_{ij}|\le T \right\}\,, \ \ T > 0\,,
\end{equation}
inside $GL(n,\Z)$.
For large $T$ one has $\#\Gamma_T \sim c_n T^{n^2-n}$, for some positive constant $c_n$.\footnote{See \cite[Corollary~2.3]{GN} and \cite[(1.14)]{DRSm} for more details on this   surprisingly difficult result.}
We now consider a series of algorithms to sample matrices in $GL(n,\Z)$.  The  most naive way to uniformly sample $\G_T$   is prohibitively slow:

\begin{center}
\fbox{\fbox{\begin{minipage}{11.5cm}
\begin{algorithmic}
\STATE {\bf \underline{Algorithm 1.}}
\vspace{.25cm}
\STATE For each $1\le i,j \le n$ sample $m_{i,j}\in \Z\cap [-T,T]$ at random.
\STATE Let $M=(m_{ij})$.
\STATE Discard and repeat if $\det(M)\neq \pm 1$, otherwise
\RETURN $M$.
\end{algorithmic}
\end{minipage}}}
\end{center}
Though we do not analyze it here, the determinant of such a randomly chosen matrix $M$ is a very large integer, and highly improbable to be $\pm 1$ as required for membership in $GL(n,\Z)$.  One minor improvement that can be made is to first check that the elements of each row (and of each column, as well) do not share a common factor, which is a necessary condition to have determinant $\pm 1$.  Nevertheless, this fails to seriously improve the extreme unlikelihood of randomly producing an integral matrix of determinant $\pm 1$.

\begin{center}
\fbox{\fbox{\begin{minipage}{11.5cm}
\begin{algorithmic}
\STATE {\bf \underline{Problem 4.}}
\vspace{.25cm}
\STATE Find a nontrivial uniform sampling algorithm which substantially speeds up Algorithm 1.
\end{algorithmic}
\end{minipage}}}
\end{center}
We note that  some computer algebra packages   include commands for generating random elements of $GL(n,\Z)$.  In addition to its command {\tt RandomSLnZ} which we shall shortly come to in Algorithm 2,
Magma's documentation includes the command {\tt RandomUnimodularMatrix} for fairly rapidly generating matrices in $GL(n,\Z)$ (not $SL(n,\Z)$ as the name indicates) having ``most entries'' inside a prescribed interval, but provides no further explanation.  Even after accounting for a typo which switches the role of the command's arguments, we found that in fact most of the entries were {\it outside} the prescribed interval (the documentation's claims notwithstanding).  Furthermore, the lattices constructed using this command appear to be much easier to attack than those generated by the closest analog considered here (Algorithm 4).     SageMath's {\tt random\_matrix} command has a {\tt unimodular} constructor (designed for teaching purposes) which does produce matrices in $GL(n,\Z)$ whose entries are bounded by a given size, but it is not as fast as other alternatives and its outputs must satisfy further constraints.  For these reasons we did not seriously examine {\tt RandomUnimodularMatrix} and {\tt random\_matrix}.

Because  Algorithm 1 is so slow, the rest of this section considers faster algorithms which do {\it not} uniformly sample $\G_T$, some coming closer than others.\footnote{Unfortunately it is prohibitively complicated here to describe particular parameter choices matching the bound in (\ref{GammaT}).} For $1\le i\neq j \le n$ let $E_{i,j}$ denote the elementary $n\times n$ matrix whose entries are all 0 aside from a 1 in the $(i,j)$-th position.  Here as elsewhere the abbreviation ``i.i.d.'' stands for ``independently identically distributed''.

\begin{equation}\label{alg2}
\text{
\fbox{\fbox{\begin{minipage}{11.5cm}
\begin{algorithmic}
\STATE {\bf \underline{Algorithm 2 (Random products of unipotents,}
\STATE {\bf  \underline{such as Magma's {\tt RandomSLNZ}).}}}
\vspace{.25cm}
\STATE {\bf Input:} a size bound $b$ and word length $\ell$.
\STATE {\bf Return:}  a random product $\g_1\cdots \g_\ell$, where each $\g_k$ is chosen i.i.d.~uniformly among all $n\times n$ matrices of the form $I_n+xE_{i,j}$, with $i\neq j$  and $x\in \Z\cap [-b,b]$.
\end{algorithmic}
\end{minipage}}}}
%
\end{equation}
 As we shall later see, the matrices produced by Algorithm 2 have a very special form, creating a cryptographic weakness.

  Algorithm 2 can be thought of as a counterpart to the LLL algorithm \cite{LLL}, which applies successive unipotent matrices and vector swaps to reduce lattices.  Although Algorithm 2 does not literally contain vector swaps, they are nevertheless present in the background because conjugates of $\gamma_j$ by permutation matrices have the same form $I_n+xE_{i,j}$ as $\gamma_k$.    In that light, the following algorithm can then be thought of as an analog of BKZ reduction \cite{BKZ}, since it utilizes block matrices of size much smaller than $n$.  Its statement involves the embedding maps $\Phi_{k_1,\ldots,k_d}:GL(d,\R)\hookrightarrow GL(n,\R)$ for size-$d$ subsets $\{k_1,\ldots,k_d\}\subset \{1,\ldots,n\}$,
\begin{equation}\label{embeddingmap}
  (\Phi_{k_1,\ldots,k_d}(h))_{i'j'} \ \ = \ \
  \left\{
    \begin{array}{ll}
      h_{ij}, & \text{if }i'=k_i\text{ and }j'=k_j \text{ for some }i,j\le d; \\
      \delta_{i'=j'}, & \hbox{otherwise\,,} \\
    \end{array}
  \right.
\end{equation}
where $h=(h_{ij})\in GL(d,\R)$.\footnote{The role of $GL(\cdot,\cdot)$ as opposed to $SL(\cdot,\cdot)$ here is again purely cosmetic.}  The image of $\Phi_{k_1,\ldots,k_d}$ is a subgroup of $GL(n,\R)$ isomorphic to $GL(d,\R)$.  (Of course we will only apply the map $\Phi_{k_1,\ldots,k_d}$ to elements of $GL(d,\Z)$.)

\begin{center}
\fbox{\fbox{\begin{minipage}{11.5cm}
\begin{algorithmic}
\STATE {\bf \underline{Algorithm 3 (Random products of smaller matrices).}}
\vspace{.25cm}
\STATE {\bf Input:}~a word length $\ell$ and  fixed  dimension $2\le d < n$ for which one can uniformly\footnote{More generally, one can consider non-uniform distributions as well.} sample $GL(d,\Z)$ matrices in a fixed box.
\STATE {\bf Return:}~a random product
  $\gamma_1\cdots \gamma_\ell$
 in which  each
  $\g_j\in GL(n,\Z)$ is a matrix of the form
  $ \Phi_{k_1,\ldots,k_d}(\g^{(d)})$,
 where $\g^{(d)}$ is   a uniformly sampled random element of  $GL(d,\Z)$ in the fixed box   mentioned above, and
 $\{k_1,\ldots,k_d\}$ is a uniformly sampled random subset of $\{1,\ldots,n\}$ containing $d$ elements.
\end{algorithmic}
\end{minipage}}}
\end{center}

We expect Algorithm 3 produces more-uniformly distributed matrices as $d$ increases.  The role of the parameter $d$ is essentially to interpolate between Algorithm 1 (which is the case $d=n$) and Algorithm 2 (which is close to the case $d=2$, but not exactly:~$\gamma^{(2)}$ need not be unipotent).

Next we turn to the following method, which among the algorithms we considered seems the best at rapidly creating uniformly-distributed entries of matrices in $GL(n,\Z)$.  This algorithm was originally suggested to us by Joseph Silverman in a slightly different form, in which more coprimality conditions needed to be checked.  It relies on the fact that an integral $n\times n$ matrix
 $M =(m_{ij})$ lies in $GL(n,\Z)$ if and only if the $n$ determinants of $(n-1)\times (n-1)$ minors
\begin{multline}\label{nminors}
  \det\(\begin{smallmatrix}
m_{22} & \cdots &  m_{2n} \\
\vdots & \ddots & \vdots \\
m_{n2} & \cdots & m_{nn}
\end{smallmatrix}\),
 \
  \det\(\begin{smallmatrix}
m_{21} & m_{23} & \cdots &  m_{2n} \\
\vdots & \vdots & \ddots & \vdots \\
m_{n1} & m_{n3} & \cdots & m_{nn}
\end{smallmatrix}\),
\ldots,
  \det\(\begin{smallmatrix}
m_{21} & \cdots &  m_{2\,n-1} \\
\vdots & \ddots & \vdots \\
m_{n1} & \cdots & m_{n\, n-1}
\end{smallmatrix}\)
\end{multline}
share no common factors.
\begin{center}
\fbox{\fbox{\begin{minipage}{11.5cm}
\begin{algorithmic}
\STATE {\bf \underline{Algorithm 4 (slight modification of a suggestion of}}
\STATE {\bf \underline{ Joseph Silverman).}}
\vspace{.15cm}
\STATE Uniformly sample random integers $m_{i,j}\in [-T,T]$, for $2\le i \le n$ and $1\le j \le n$,
until the $n$ determinants in (\ref{nminors}) share no common factor.
\STATE Use the
euclidean algorithm to find integers $m_{11},\ldots,m_{1n}$ such that
$\det((m_{ij}))=\pm 1$, the sign chosen uniformly at random.
\STATE Use least-squares to find the
linear combination $\sum_{i\ge 2}^n c_i [m_{i1}\cdots m_{in}]$ closest to $[m_{11} \cdots m_{1n}]$,
and let $\widetilde{c_i}$ denote an integer nearest to $c_i$.
\STATE {\bf Return:}   the matrix $M$ whose top row is
$$[m_{11} \cdots m_{1n}]\, - \, \sum_{i\ge 2}^n  \widetilde{c_i} [m_{i1}\cdots m_{in}]$$
and whose $i$-th row (for $i\ge 2$) is $[m_{i1}\cdots m_{in}]$.
\end{algorithmic}
\end{minipage}}}
%
\end{center}

\noindent
{\bf Remarks on Algorithm 4:}
The $n$ large integers in (\ref{nminors}) are  unlikely to share a common factor:~for example, the most probable common factor is 2, which happens only with probability $\approx 2^{-n}$.  Obviously the top row of $M$ is chosen differently than the others, and its size is different as well since it typically has entries larger than size $T$ -- this is because the euclidean algorithm can produce large coefficients (as the minors in (\ref{nminors}) are themselves so enormous).  Also, it is likely that the first two or three minors will already be coprime, and hence that most  of the entries in   $[m_{11}\,m_{12}\,\cdots\,m_{1n}]$ will vanish.   The use of rounding and least-squares  cuts down this size and further randomizes the top row, while keeping the determinant equal to one.

One could instead try a different method to find an integral combination of the bottom $n-1$ rows closer to the initial guess for the top row.  One extreme possibility involves appealing to the Closest Vector Problem (CVP) itself, which is thought to be very difficult.
 We found Algorithm 4 gave good randomness properties in that nearly all of the matrix is equidistributed, and it is fairly fast to execute.  In comparison, we will see that using Algorithm 2 requires many matrix multiplications to achieve random entries of a similar size, which are not as well distributed anyhow.

The following algorithm is folklore and has appeared in various guises in many references (for example \cite{CHKPeik}, which uses Gaussian sampling and has provable hardness guarantees,\footnote{It should be mentioned that provable guarantees were earlier established in \cite{Ajtai,AlPeik,MiccPeik} when one generates both the lattice together with a basis at random from a family.  Here our emphasis is on a fixed, given lattice.} though not necessarily for Problem 3).  As we shall see just below, it shares some similarities with Algorithm 4.
\begin{center}
\fbox{\fbox{\begin{minipage}{11.5cm}
\begin{algorithmic}
\STATE {\bf \underline{Algorithm 5 (via Hermite Normal Form).}}
\vspace{.15cm}
\STATE Create a uniformly
distributed $m\times n$ matrix $B$, with $m\ge n$ and entries uniformly chosen in $\Z\cap [-T,T]$.
\STATE Decompose $B$ in a Hermite normal form
$B=UM$, where $M\in GL(n,\Z)$ and $U=(u_{ij})$ has no nonzero entries with $i<j$.
\STATE {\bf Return:} $M$.
\end{algorithmic}
\end{minipage}}}
\end{center}

\noindent
{\bf A surprising connection between Algorithms 4 and 5:}
Even though Algorithms 4 and 5 appear to be very different, they are actually extremely similar (in fact, arguably nearly identical) in practice.  Algorithms for Hermite Normal Form (such as {\tt HermiteDecomposition} in Mathematica) proceed by building the matrix $M$ directly out of the rows of $B$ whenever possible.  For example, it is frequently the case that the first $n-1$ rows of $U$ agree with those of the identity matrix $I_n$, or at least differ only very slightly; in other words, the
  first $n-1$ rows of $B$ and $M$ are expected to coincide or nearly coincide.\footnote{In our experiments, for example, the top $n-2$ rows agreed most of the time for $m=n\ge 10$.}
Also, the last row of $M$ is an integral combination of the first $n$ rows of $B$.  In contrast with Algorithm 4 this last combination, however, is mainly determined by arithmetic considerations, and in particular depends on the $n$-th row of $B$; thus more  random information is used  than in Algorithm 4, which uses only $n^2-n$ random integers instead of the $n^2$ here.\footnote{Note the order of magnitude of the set $\Gamma_T$ from (\ref{GammaT}) is $T^{n^2-n}$, naturally matching the $n^2-n$  random integers picked in Algorithm 4.}

 To summarize, in fairly typical cases both Algorithms 4 and 5 populate the matrix $M$ by first generating all but one row uniformly at random, and then using integral combinations to create a final row having relatively small entries.  The practical distinction is essentially how this final row is created, which utilizes further random information in Algorithm 5 but not in Algorithm 4.
The final row also appears to be typically smaller (that is, closer to fitting in the box defined in (\ref{GammaT})) when using Algorithm 4 than when using Algorithm 5; consequently, we did not perform any experiments with Algorithm 5.

Note that the Hermite decomposition as stated above is not unique, since there are lower triangular matrices in $GL(n,\Z)$.  Thus there can be no immediate guarantee on the entry sizes of $M$ unless this ambiguity is resolved.  Algorithm 5 can be thought of  as a $p$-adic analog of the following method of producing random rotations in $O(n)$:~apply the Gram-Schmidt orthogonalization process to a matrix chosen according to a  probability density function (e.g., Gaussian) which is invariant under multiplication by $O(n)$.\\

\noindent
{\bf Remarks on an Algorithm in \cite{rivin}:}~Igor Rivin makes the   proposal in \cite[\S6.1]{rivin} to generate matrices in $GL(n,\Z)$ by applying complete lattice basis  reduction to a  basis of $\R^n$ chosen inside a large ball.  Let $B\in GL(n,\R)$ denote the $n\times n$ matrix whose rows consist of this  basis. Complete lattice reduction produces a random element $\gamma \in GL(n,\Z)$ of constrained size for which $\gamma B$ lies in a fixed fundamental domain for $GL(n,\Z)\backslash GL(n,\R)$.

This procedure is extremely slow, since complete lattice reduction is impractical in large dimensions.  Rivin thus considers instead using  weaker lattice basis reduction methods (such as LLL \cite{LLL})  to speed this up, but at the cost of less-uniform distributions.  For example, the results of  LLL are  thought to be skewed towards certain favored outputs avoiding ``dark bases'' \cite{darkbases}.  Since our interest in
generating random bases is to see how long incomplete lattice reduction takes on them, the use of   lattice reduction to itself make the basis itself is too slow for our purposes (hence we did not consider this algorithm in our experiments).

\section{Experiments on recognizing $\Z^n$}\label{sec:experiments}

In this section we report on attempts to solve  Problem 2b on instances of  matrices $M$ generated using some of the algorithms  from Section~\ref{sec:randomGLNZ} for sampling  $GL(n,\Z)$.  We first note that Geissler and Smart \cite{GeisslerSmart} reported on attempts to solve Problem 2b on NTRU lattices using LLL \cite{LLL} (as well as their own modification, for which they report  up to a factor of four speedup), and concluded from lattice reduction heuristics that LLL itself is insufficient for NTRU instances with dimensions and matrix entry size far smaller than those considered in (\ref{NTRUsecurityb}) below (see Appendix~\ref{app:NTRU}).  Nevertheless LLL performs fairly well on rotations of the $\Z^n$ lattice  as  compared to on a random lattice, which is not unexpected since the latter  has shortest vector on the order of $\sqrt{n}$ (as opposed to 1 for rotations of the $\Z^n$ lattice).  Given that LLL typically outperforms its provable guarantees, it is not surprising it is fairly effective on Problem 2b.

 Our main emphasis is that LLL and BKZ perform better on certain distributions with respect to  Problem 2b than on others.
  Instead of LLL alone, we try the following:
\begin{equation}\label{testingprocedure}
\text{
\fbox{\fbox{\begin{minipage}{11.5cm}
 {\bf \underline{Procedure to test matrix generation algorithms}\newline \underline{with Problem 2b.}}
\begin{enumerate}
  \item In Magma, apply LLL or Nguyen-Stehl\'e's L2 lattice basis reduction algorithm \cite{NS} to the Gram matrix $G=MM^t$, then
  \item apply BKZ with incrementally-increasing block sizes $3,4,$ and $5.$
  \item Success is measured by whether or not the output basis vectors all have norm equal to 1 (in which case they span a rotation of the $\Z^n$ lattice).
\end{enumerate}
\end{minipage}}}}
\end{equation}
We chose to use Magma's built-in lattice basis reduction routines, partly because
of slow running times with other implementations (such as fplll in SageMath) on matrices with very large integer entries.
 In step 2 one can of course continue further with block sizes larger than 5, but we fixed this as a stopping point in order to be systematic.

Our main finding is that Algorithm 2 in \secref{sec:randomGLNZ} (as implemented in Magma's {\tt RandomSLnZ}) is insecure  for generating hard instances of Problem 2b.  Algorithms 3, 4, and 5 fare much better.  It is not surprising that Algorithm 5 (and the nearly-equivalent Algorithm 4)  give harder instances, since there are provable guarantees attached to Algorithm 5 in a different context \cite{CHKPeik};  there is a serious  difference between these and Algorithm 2 described below and in Appendices~\ref{app:alg3} and~\ref{app:alg4}.

\subsection{Experiments with Algorithm 2 (Magma's {\tt RandomSLnZ} command)}

We begin with some comments on entropy and generating random products with a constrained number of bits.
To mimic random elements of $GL(n,\Z)$, one may desire that the product matrix  has as many nonzero entries as possible per random bit.  For this reason, our experiments set the parameter $b=1$ in Algorithm 2 in order to take longer products (thereby further increasing the number of nonzero entries of the matrix), while keeping the number of random bits constant.  When the product length is less than $n$, one expects to have rows or columns of the product matrix which are unchanged by the successive matrix multiplications.  (This much less likely to be  the case for the Gram matrices, however.)

Thus each random factor has at most a single nonzero off-diagonal entry, which is $\pm1 $. It is prohibitive to pack in as many random bits as the total number of entries  this way, since   multiplication of large matrices is slow. As an extreme example, as part of a comparison with the last row of (\ref{NTRUsecuritya}) we generated a random matrix in $GL(1486,\Z)$ using products of length 55,000, again with $b=1$.  Generating the product alone took about half a day. Its row lengths were between $2^{14}$ and $2^{20}$ in size.  For comparison, an NTRU matrix with similar row lengths (as in Table~\ref{NTRUsecuritya}) uses 8,173 random bits.  The comparison with NTRU is made here simply because concrete bit-strengths have been asserted for NTRU lattices; this is why we took the particular values of $n$ in (\ref{NTRUsecurityb}) (see Appendix~\ref{app:NTRU} for more details).  One might hypothesize that having more random bits in the matrix makes solving Problem 2b more difficult, but as we shall see this  in fact turns out   to not always be the case:~the structure of the matrix plays a very important role, and the product structure from Algorithm 2 seems to be a contributing weakness.  In particular, the larger the value of the parameter $b$, the more unusual properties the product matrix possesses.


\begin{equation}\label{NTRUsecurityb}
\gathered
\text{
{\bf{Successful experiments on large  lattices}}}
\\
\text{
\begin{tabular}{|c|c|c|c|c|}
\hline
  $n=\dim(\L)$ & estimated bit-hardness &  range of vector lengths  & product  \\
  &  for corresp. NTRU (\ref{NTRUsecuritya}) & (in bits) &  length \\
\hline
 886 & 128 & from $25$ to $32$& 55,000\\
 1486 & 256 & from $14$ to $20$ & 55,000\\
\hline
\end{tabular}
}
\endgathered
\end{equation}
  From the success of our trials one immediately sees the Lenstra-Silverberg Problem 2b is fairly easy for matrices $M$ generated by  Magma's {\tt RandomSLnZ} command.  (Of course it is well known to be impossible to solve Problem 2b using LLL or BKZ with small block sizes on NTRU matrices of the comparable size listed in (\ref{NTRUsecurityb}) and (\ref{NTRUsecuritya}), or even those much smaller.)

\subsection{Experiments with Algorithm 3 (random $GL(d,\Z)$ matrices)}

Next we consider matrices generated by Algorithm 3 (random $GL(d,\Z)$'s), and find that for small $d$ they are also cryptographically weak for the Lenstra-Silverberg problem, but stronger than those generated by Algorithm 2.  Furthermore, we see their strength increases with increasing $d$.

 The tables in Appendix~\ref{app:alg3} list the outcomes of several  experiments   attacking instances of Problem 2b for matrices $M$ generated by Algorithm 3.
 One sees the dramatic effect of the product length $\ell$.  For example, if $\ell$ is too short there may be rows and columns of the matrix not touched by the individual   multiplications by the embedded random $d\times d$ matrices; if $\ell$ is too long, the matrix entries become large and lattice basis reduction becomes difficult.

\subsection{Experiments with Algorithm 4}

Finally, we turn to the opposite extreme of random elements of $GL(n,\Z)$ generated by Algorithm 4, in which the bottom $n-1$ rows are uniformly distributed among entries in the range $[-T,T]$.  Here we were able to solve Problem 2b with instances having $n=100$, even with entry sizes up to $T=50$ (again, using the testing procedure in (\ref{testingprocedure})).  However, none of our experiments with $n\ge 110$ were successful at all, even with $T=1$ (i.e., all entries below the top row are $-1$, $0$, or $1$).  See the tables in Appendix~\ref{app:alg4} for more details.

\section{Random basis generation in the DRS NIST Post-Quantum Cryptography competition submission}\label{sec:DRS}

In \cite[\S 5.1.2]{AEN} some examples of methods for generating random lattice bases are described, which are closely related to Algorithms 2, 3, and 5.  The authors reported their experiments on those methods resulted in similar outcomes in practice.  Our experiments, however, do show a difference (as was explained in Section~\ref{sec:experiments}).

In this section we wish to make further comments about one method   highlighted in \cite{AEN}, which is from the DRS NIST Post-Quantum competition submission \cite[\S 2.2]{DRSc}.  Random elements of $GL(n,\Z)$ there are constructed as products of length $2R+1$ of the form
\begin{equation}\label{DRS1}
 P_1 \gamma_1 P_2 \gamma_2 P_3 \gamma_3\cdots P_R \gamma_R P_{R+1}\,,
\end{equation}
where $P_1,\ldots,P_{R+1}$ are   chosen uniformly at random among permutation matrices in $GL(n,\Z)$ and $\gamma_1,\ldots,\gamma_R$ are elements in $SL(n,\Z)$ produced by the following random process.  Let $A_+=\ttwo 1112$ and $A_{-}=\ttwo 1{-1}{-1}2$.  Then each $\gamma_i$ is a block diagonal matrix with $\frac n2$  $2\times 2$ entries chosen uniformly at random from $\{A_+,A_{-}\}$.  This construction has some similarities with Algorithm 3 for $d=2$, but note that here many of the $SL(2)$ matrices commute (being diagonal blocks of the same matrix).  In fact, since  $A_+$ is conjugate by $\ttwo 100{-1}$ to $A_{-}$    one may replace each $\gamma_j$ with the block diagonal matrix
$$D \ \ = \ \  \operatorname{diag}(A_+,A_+,\ldots,A_+)\,,$$
 at the cost of allowing the $P_i$'s to be signed permutation matrices.  Alternatively, by rearranging the permutation matrices and applying an extra rotation on the right, Problem 2b on matrices of the form (\ref{DRS1}) is equivalent to it on  products of the form
\begin{equation}\label{DRS2}
  M \ \ = \ \ M_1M_2\cdots M_R\,,
\end{equation}
in which each $M_i$ is conjugate of
$D$
   by a random signed permutation matrix.

Since Algorithm 3 with $d=2$ performed relatively weakly in the experiments of Section~\ref{sec:experiments},  we suspect  Problem 2b is relatively easy to solve on matrices generated using (\ref{DRS1}) (as compared to those, say, generated using Algorithm 4).  The experiments described below bear this out.  (All of our   remaining comments in this section pertain solely to   (\ref{DRS1}) in the context of Problem 2b, and not to any other aspect of \cite{DRSc}.)

The parameters listed in \cite[\S3.2]{DRSc} assert 128-bit security for their scheme when $(n,R)=(912,24)$, 192-bit security when $(n,R)=(1160,24)$, and 256-bit security when $(n,R)=(1518,24)$.   Our main finding is that the testing procedure (\ref{testingprocedure}) was able to recover $M$ chosen with the 256-bit security parameters in 708 hours of running time.  We could also recover $M$ chosen with the 192-bit security parameters in 222 hours of running time but (as we describe below) could not fully recover $M$ with the 128-bit security parameters.

\begin{figure}[ht!]
\includegraphics[width=\linewidth]{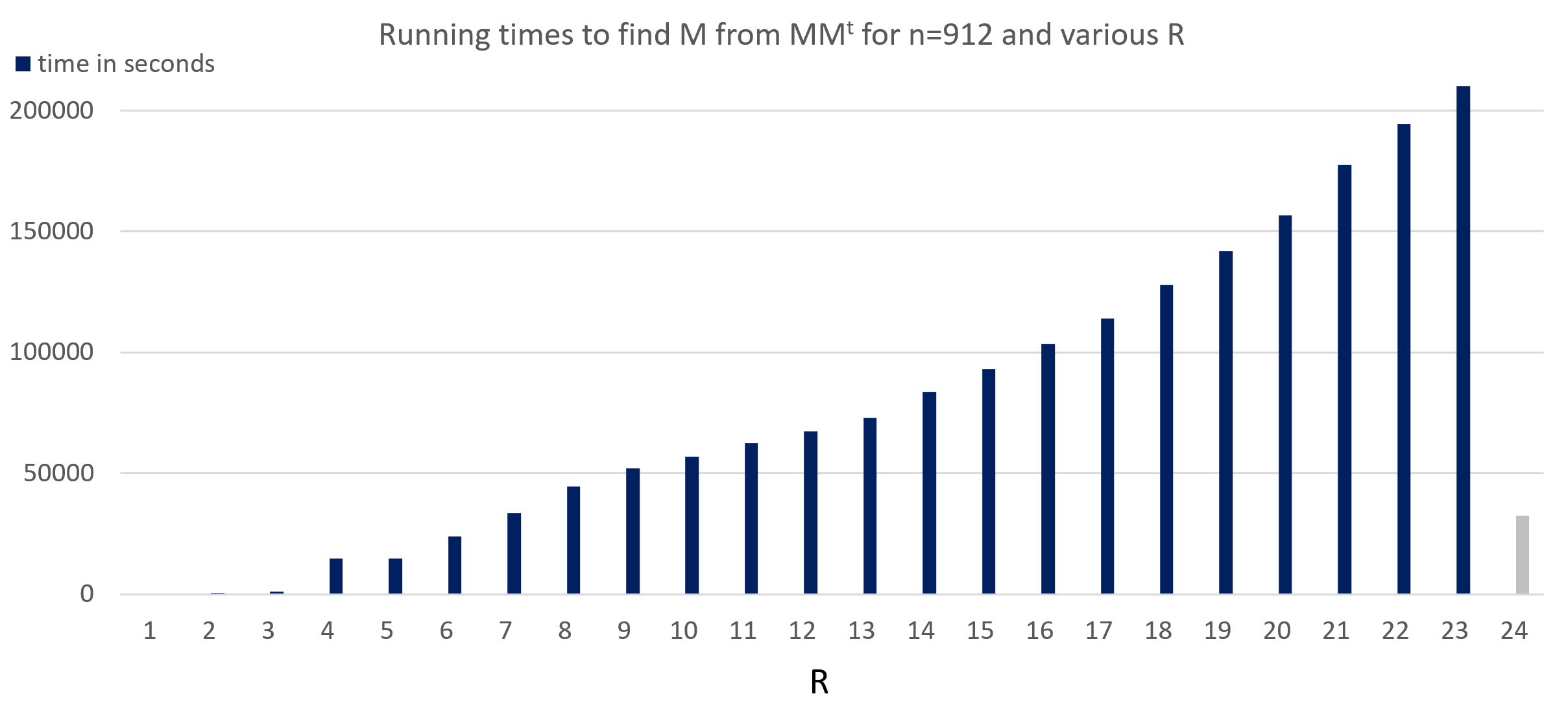}
\caption{
We experimentally tried to solve Problem 2b on instances generated by  the random basis construction from the DRS NIST submission \cite[\S2.2]{DRSc}, using its suggested parameters $(n,R)=(912,24)$ for 128-bit security.  This failed with $n=912$ and $R=24$ itself (the gray bar on the right), but was successful for $n=912$ and  $1\le R \le 23$.  We were able to solve all cases for $R\le 22$ in less than 60 hours using LLL alone, and the $R=23$ case in slightly more time using the procedure in (\ref{testingprocedure}).
We conclude that  method of random basis generation in the DRS digital signature scheme is insecure with the recommended parameter setting $(n,R)=(912,24)$, at least for Problem 2b.  Times are shown for runs on a Dell PowerEdge R740xd server equipped with two Intel Xeon Silver 4114 2.2GHz processors and 256GB RAM.
 \label{fig:DRSplot}}
\end{figure}

 The testing procedure (\ref{testingprocedure}) also easily solves Problem 2b when $n$ or $R$ are smaller yet still relatively large.  For example, it took  roughly an hour to recover $M$ from $MM^t$ when $(n,R)=(180,24)$ using BKZ with block sizes up to 26.
In Figure~\ref{fig:DRSplot} we show the results of several experiments for the parameter choice of $n=912$ and increasing values of $R$ up to the recommended choice of $R=24$ for 128-bit security.   The results were strikingly successful, in that  each trial for $R\le 22$  successfully recovered $M$ from $MM^t$ using only LLL (without requiring BKZ). We additionally tried $R=23$ and {\it nearly} recovered $M$ using LLL this way:~the longest vector in the LLL output had length $\sqrt{7}$, and subsequently applying BKZ reduction with block size 3 for less than five minutes then fully recovered $M$.  However, we were unsuccessful in the  $R=24$ case suggested in \cite{DRSc}.

\begin{figure}[h!]
\includegraphics[width=5in]{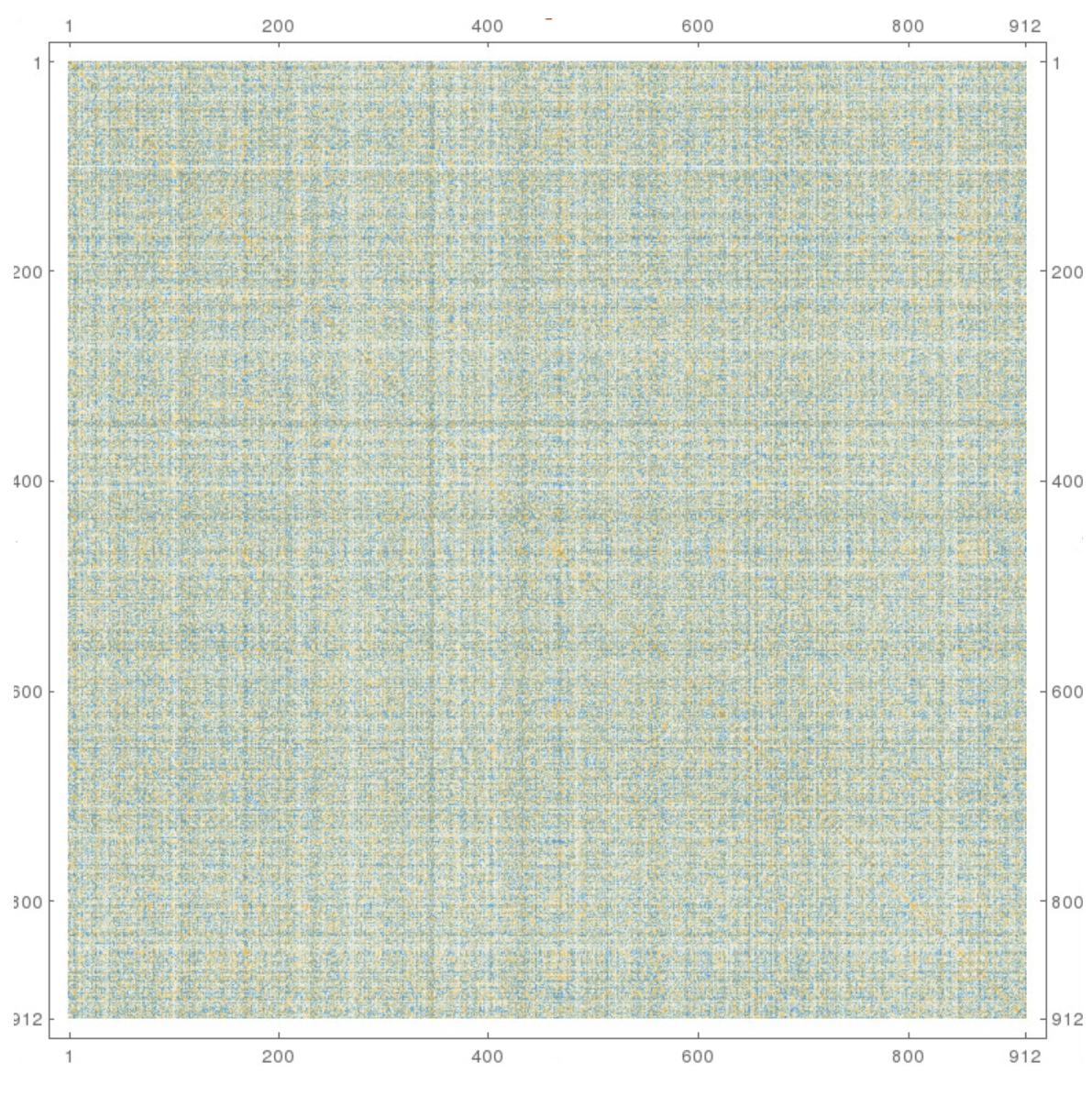}
\caption{Mathematica's {\tt MatrixPlot} command  displays the nonzero entries of Gram matrix $MM^t$ as darkened pixels, where $M$ was generated according to (\ref{DRS1}) with recommended parameters $n=912$ and $R=24$ from \cite{DRSc}.  Similarly banded plots arise when $M$ is generated using Algorithm 3 with $d=2$.  In contrast, Gram matrices generated by   Algorithm 4 have a (provably) far more uniform structure.   \label{fig:sofa}}
\end{figure}

Again, these results are {\em only} for Problem 2b applied to the random basis construction used in the DRS digital signature scheme \cite{DRSc}; nevertheless, this may indicate a weakness in the digital signature scheme as well.
Somewhat counterintuitively, our experiments for fixed values of the product length parameter $R$ sometimes fared better for {\em larger} values of $n$.  For example, we were   successful with $(n,R)=(912,22)$ despite not being successful for $(n,R)=(200,22)$, and we were successful with $(n,R)=(1160,24)$ and $(1518,24)$ despite not being successful for $(n,R)=(912,24)$.  Our explanation is that as $n$ grows   there may be a weakness in that it is hard to randomly fill out the full matrix $M$ (a similar phenomenon occurs in Algorithms 2 and 3 for small $\ell$). Indeed, matrices of the form (\ref{DRS1}) seem to have a very special form:~Figure~\ref{fig:sofa} shows the entry sizes in $MM^t$ have a  banded structure.

\section{Conclusions}\label{sec:conclusions}

We have considered the role of generating random elements in $GL(n,\Z)$ in the difficulty of lattice problems, and have found that it can have a profound influence.  Concretely, Magma's {\tt RandomSLnZ} command (Algorithm 2) gives easy instances of Lenstra-Silverberg's ``Recognizing $\Z^n$ Decision'' Problem 2b from (\ref{LSQGabnew}).  We were able to successfully attack lattices of dimension up to 1,486, which are in some measurable ways comparable to NTRU lattices having claimed 256-bit quantum security.  On the other hand,  using  the apparently stronger methods  of Algorithms 3 and 4  make Problem 2b much more difficult to solve (as expected).

We would thus recommend not using  Algorithm 2 in generating random bases for cryptographic applications.  We also recommend not using the random basis algorithm from the NIST Post-Quantum Competition submission DRS \cite{DRSc}, because we  were similarly able to solve Problem 2b on instances of its random basis generation method with its recommend  parameters for 256-bit security.

We have not fully understood the weaknesses of these algorithms.  It seems plausible that the failure to quickly fill out the matrix entries in a uniform way  is at least partly to blame, since many do not get sufficiently randomized.  The construction of Algorithm 2 in some sense reverses the steps of an {\tt LLL} basis reduction, which might explain why LLL is particularly effective against it.  More generally one might expect the block sizes in Algorithm 3 to be related to the block sizes in the {\tt BKZ} algorithm.  It is natural from this point of view to expect Algorithms 4 and 5 to be the strongest lattice basis generation algorithms considered in this paper, consistent with the results of our experiments.

\appendix

\section{Experiments with Algorithm 3 (random $GL(d,\Z)$ matrices)}\label{app:alg3}

Below we list tables of the experimental results mentioned in Section~\ref{sec:experiments} on Algorithm 3, performed using the testing procedure (\ref{testingprocedure}).

\begin{center}

\begin{tabular}{|r|r|r|r||r|r||c|}
  \hline
$n$ & $d$ & $T$ & $\ell$ & shortest row   & longest row  & found $M$? \\
    &       &   &     & length (in bits) & length (in bits)  & \\
\hline
\hline
200&2&1&4000& 6.03607& 12.7988 & $\times$\\
\hline
200&2&2&1500&1.29248& 18.5329 & $\checkmark$ \\
200&2&2&2000&7.86583& 22.2151 & $\times$ \\
\hline
200&2&3&1000&0.5& 27.0875 & $\times$ \\
200&2&3&2000&23.521& 41.5678 & $\times$ \\
\hline
200&2&10&500&2.04373& 38.7179 & $\checkmark$ \\
200&2&10&700&7.943& 49.0346 & $\times$ \\
\hline
\hline
200&3&1&1000&2.04373& 11.3283 & $\checkmark$ \\
200&3&1&1500&7.66619& 17.1312 & $\times$ \\
200&3&1&2000&13.0661& 20.8768 & $\times$ \\
\hline
200&3&2&500&3.27729& 18.4087 & $\checkmark$ \\
200&3&2&600&4.89232& 24.111 & $\times$ \\
200&3&2&1000&13.0585& 34.0625 & $\times$ \\
\hline\hline
200&4&1&500&3.66096& 12.2277 & $\checkmark$ \\
\hline
200&4&2&300&0.5& 24.2424 & $\checkmark$ \\
200&4&2&400&1.79248& 26.6452 & $\times$ \\
\hline
\end{tabular}
\end{center}
\noindent {\bf key:} $n$=lattice dimension, $d$=size of smaller embedded matrices, $T$=bound on embedded matrix entries, $\ell$=length of the product of smaller matrices.

\begin{center}
\begin{tabular}{|r|r|r|r||r|r||c|}
  \hline
$n$ & $d$ & $T$ & $\ell$ & shortest row   & longest row  & found $M$? \\
    &       &   &     & length (in bits) & length (in bits)  & \\
\hline
\hline
500&2&1&4000&0.& 5.90085 & $\checkmark$ \\
500&2&1&8000&3.41009& 10.7467 & $\checkmark$ \\
500&2&1&10000&7.08508& 12.7447 & $\checkmark$ \\
500&2&1&15000&12.6617& 18.5326 & $\checkmark$ \\
500&2&1&20000&18.0246& 24.5732 & $\times$ \\
\hline
500&2&2&4000&4.21731& 18.587 & $\checkmark$ \\
500&2&2&6000&12.3467& 28.7882 & $\times$ \\
500&2&2&8000&18.87& 35.7267 & $\times$ \\
500&2&2&10000&28.5508& 45.8028 & $\times$ \\
\hline
500&2&3&2000&0.& 19.0752 & $\checkmark$ \\
500&2&3&3000&7.38752& 32.9895 & $\checkmark$ \\
500&2&3&4000&16.9325& 40.9656 & $\times$ \\
\hline
500&2&10&1000&0.& 30.3755 & $\checkmark$ \\
500&2&10&2000&11.9964& 61.5006 & $\times$ \\
\hline\hline
500&3&1&1000&0.& 5.39761 & $\checkmark$ \\
500&3&1&2000&1.29248& 9.164 & $\checkmark$ \\
500&3&1&3000&2.37744& 13.9903 & $\checkmark$ \\
500&3&1&4000&8.43829& 17.4593 & $\checkmark$ \\
500&3&1&5000&14.1789& 21.528 & $\checkmark$ \\
500&3&1&6000&18.3878& 25.2578 & $\times$ \\
500&3&1&7000&20.5646& 29.287 & $\times$ \\
\hline
500&3&2&1000&0.& 15.551 & $\checkmark$ \\
500&3&2&2000&3.24593& 33.0945 & $\checkmark$ \\
500&3&2&3000&23.5966& 43.7986 & $\times$ \\
\hline
500&3&3&1000&0.& 28.1575 & $\checkmark$ \\
500&3&3&2000&16.6455& 53.1806 & $\times$\\
500&3&3&3000&41.3371& 83.9486 & $\times$ \\
\hline\hline
500&4&1&1000&0.& 9.85319 & $\checkmark$ \\
500&4&1&2000&8.11356& 18.9434 & $\checkmark$ \\
500&4&1&3000&19.1019& 26.9836 & $\checkmark$ \\
500&4&1&4000&24.4869& 35.6328 & $\times$ \\
500&4&1&5000&26.6804& 44.3982 & $\times$ \\
500&4&1&6000&40.5944& 53.3654 & $\times$\\
\hline
500&4&2&1000&6.29272& 33.4373 & $\checkmark$ \\
500&4&2&2000&33.6181& 63.3469 & $\times$ \\
  \hline
\end{tabular}
\end{center}
\noindent {\bf key:} $n$=lattice dimension, $d$=size of smaller embedded matrices, $T$=bound on embedded matrix entries, $\ell$=length of the product of smaller matrices.

\begin{center}
\begin{tabular}{|r|r|r|r||r|r||c|}
  \hline
$n$ & $d$ & $T$ & $\ell$ & shortest row   & longest row  & found $M$? \\
    &       &   &     & length (in bits) & length (in bits)  & \\
\hline
\hline
886&2&1&3000 & 0& 3.49434 & $\checkmark$\\
886&2&1&4000 & 0& 3.80735 & $\checkmark$\\
886&2&1&5000 & 0& 4.40207 & $\checkmark$\\
886&2&1&6000 &  0& 5.30459& $\checkmark$\\
886&2&1&7000 & 0& 6.16923 &$\checkmark$\\
886&2&1&8000 & 0& 6.90754 &$\checkmark$\\
886&2&1&9000 & 1& 7.58371 &$\checkmark$\\
886&2&1&10000& 2.37744& 8.05954& $\checkmark$\\
886&2&1&15000&5.46942& 11.2176 &$\checkmark$\\
886&2&1&20000&8.6594& 14.5837 &$\checkmark$\\
886&2&1&25000&10.884& 18.035 &$\checkmark$\\
886&2&1&30000&15.0082& 21.0333 &$\checkmark$\\
886&2&1&35000&17.6964& 24.8408 &$\checkmark$\\
886&2&1&40000&20.7706& 28.3888 &$\checkmark$\\
886&2&1&45000&24.484& 30.6745 &$\checkmark$\\
886&2&1&50000&25.7401& 34.0742& $\times$\\
\hline
\end{tabular}
\end{center}
\noindent {\bf key:} $n$=lattice dimension, $d$=size of smaller embedded matrices, $T$=bound on embedded matrix entries, $\ell$=length of the product of smaller matrices.

\subsection*{Comments}

Each sequence of experiments (for fixed values of $n$, $d$, and $T$) eventually fails when $\ell$ is sufficiently large.  For $\ell$ too small the random product will not involve all the rows and columns of the matrix, meaning that the dimension of the lattice problem is effectively reduced to a smaller value of $n$, so the most interesting cases are for intermediate values of $\ell$ (e.g., $10000\le \ell \le 50000$ in this last table).  There is some correlation between a successful trial and having a short vector in $M$ (the fifth column), especially in the trials for $n=200$.  For $n=500$ one sees more successful trials with longer shortest rows, especially as $d$ (and to a lesser extent, $T$) increase.  Note that each entry in these tables corresponds to a single experiment; we did not  attempt to average over several experiments since we wanted to report on the range of the row lengths.

We did not take values of $d>4$, since it is difficult to use Algorithm 1 to generate larger random elements of $GL(d,\Z)$.

The table for $n=886$ is in some sense an elaboration of the middle entry of (\ref{NTRUsecurityb}), the difference being that the latter uses unipotents (instead of embedded $GL(2,\Z)$ matrices).

\section{Experiments with Algorithm 4}\label{app:alg4}
Below we list tables of the experiments mentioned in Section~\ref{sec:experiments} on Algorithm 4, performed using the testing procedure (\ref{testingprocedure}).

\begin{center}
\begin{tabular}{|c|c||c|c||c|}
  \hline
$n$ & $T$  & shortest row   & longest row  & found $M$? \\
    &        & length (in bits) & length (in bits)  & \\
\hline
\hline
100&1& 2.91645& 4.65757& $\checkmark$\\
100&3& 4.14501& 5.81034& $\checkmark$\\
100&4& 4.50141& 6.20496& $\checkmark$\\
100&10& 5.64183& 7.15018& $\checkmark$\\
100&50& 7.99332& 9.77546& $\checkmark$\\
\hline
100&1& 2.91645& 4.65757& $\checkmark$\\
110&1& 2.98864& 4.54902& $\times$\\
120&1& 3.03304& 4.77441& $\times$\\
125&1& 3.09491& 4.93979& $\times$\\
150&1& 3.12396& 5.09738& $\times$\\
\hline
200&1& 3.42899& 5.32597& $\times$\\
200&2& 4.23584& 6.42421& $\times$\\
200&3& 4.72766& 6.82899& $\times$\\
200&4& 5.06529& 7.41803& $\times$\\
\hline
\end{tabular}
\end{center}
\noindent {\bf key:} $n$=lattice dimension,  $T$=bound on  matrix entries in bottom $n-1$ rows.

\subsection*{Comments}

In general, matrices in $GL(n,\Z)$ with large entries have very small determinants ($\pm 1$) relative to their overall entry size, so they are already very close to singular matrices.  However, the size of the rank of nearby matrices is important.
The matrices produced by Algorithm 4 are   perturbations of   matrices having rank $n-1$ (which is as large as possible for singular $n\times n$ matrices).  In contrast,
one numerically sees that matrices produced by Algorithm 2 are instead nearly rank-one matrices (i.e., up to a small overall perturbation relative to the size of the entries).      We expect Algorithm 3's matrices, which are produced by taking products of random $GL(d,\Z)$ matrices, have intermediate behavior (but  have not systematically analyzed this).

A related fact is that matrices produced by Algorithm 2 frequently have a very large row or column (if $b$ is sufficiently large) -- typically coming from the first or last factor in the matrix multiplication, respectively.  That serves as a possible hint to recover the spelling of the word in the random product, along the lines of the length-based attack in \cite[\S4]{BMV}.  However, we were unable to turn this into a direct, general attack.  For example, it is unclear what to do when the value of $x\in\Z\cap [-b,b]$ is small, say in the regime that $b\le \ell$.  (The situation is clearer when $b$ is extremely large relative to $\ell$, in which case we expect a  bias effect in random words similar to underlying device used in \cite[\S4]{BMV}.)

\section{A reference point for the bit-strength  of lattice problems:~NTRU}\label{app:NTRU}

In this appendix we give some information about how we measured when product lengths in Algorithms 2 and 3 were sufficiently long enough to ensure Gram matrix entries have an appropriately large size.  The security of lattices against LLL and BKZ is an active area in which no general consensus has been reached despite many competing suggestions (reflecting its underlying notoriously complicated difficulty).

 One type of lattice for which  bit strengths have been suggested are NTRU lattices.  We mention this as an attempt to quantify the notion that lattice problems in high dimensions are hard, as well as to provide a point of comparison --- though there are of course many differences between NTRU lattices and rotations of the $\Z^n$ lattice (we don't say anything about the security of NTRU itself).

  NTRU matrices have the form
\begin{equation}\label{NTRU1}
  \ttwo{I_{n/2}}{X}{0}{qI_{n/2}}
\end{equation}
with $n$ even, $q$ an integer greater than one, and $X$  randomly chosen from a certain distribution among all integral matrices of the form
\begin{equation}\label{NTRU2}
  X \ \ = \ \ \(
\begin{smallmatrix}
x_1 & x_2 & x_3 & \cdots & x_{n/2-1} & x_{n/2} \\
x_2 & x_3 & x_4 & \cdots & x_{n/2} & x_{1} \\
\vdots & \vdots & \vdots & \ddots & \vdots & \vdots \\
x_{n/2} & x_1 & x_2 & \cdots & x_{n/2-2} & x_{n/2-1} \\
\end{smallmatrix}
\), \ |x_j|\,\le \,\f q2\,.
\end{equation}
The rows of an NTRU matrix span an ``NTRU lattice'' $\Lambda\subset\R^n$.  In \cite{NISTRU} and in earlier NIST Post-Quantum Cryptography submissions the following quantum bit security is suggested for NTRU with the following parameters:

\begin{equation}\label{NTRUsecuritya}
\text{
\begin{tabular}{|c|c|c|}
\hline
$q$ & $n=\dim(\L)$ & estimated quantum security (in bits)\\
\hline
2048 & 886 & 128 \\
2048 & 1486 & 256 \\
\hline
\end{tabular}
}
\end{equation}
These estimates are not directly relevant to the lattice bases we examine, which have different determinants and a very different structure.  Nevertheless, they are consistent
with the general expectation that lattice problems in dimensions 500 or more (and especially 1,000 or more) become cryptographically difficult.

The choice of length $\ell$ in the experiments in (\ref{NTRUsecurityb}) was determined as follows.  The vector lengths of the rows in the NTRU matrix (\ref{NTRU1}) are either roughly $\sqrt{\f n2}\f q2$ (for the first $n/2$ rows), or exactly $q$ (for the last $n/2$ rows).  We took $\ell$ large enough so that the resulting product had comparable row lengths, and made sure to use at least as many  random bits as go into constructing an NTRU lattice (which is $\frac{n}{2}\log_2(q)$).


\begin{thebibliography}{99}

\bibitem{Ajtai} Miklos Ajtai, {\it Generating Hard Instances of the Short Basis Problem}, International Colloquium on Automata, Languages, and Programming (ICALP 1999), Springer Lecture Notes in Computer Science {\bf 1644}, pp. 1--9.

\bibitem{AlPeik} Joel Alwen and Chris Peikert, {\it Generating Shorter Bases for Hard Random Lattices}, Theory of Computing Systems {\bf 48} (2011), 535--553.

\bibitem{AEN} Yoshinori Aono, Thomas Espitau, and Phong Q. Nguyen, {\it Random Lattices:~Theory And Practice}, preprint. \url{https://espitau.github.io/bin/random_lattice.pdf}

\bibitem{BMV} Evgeni Begelfor, Stephen D. Miller, and Ramarathnam Venkatesan, {\it Non-abelian analogs of lattice rounding},
    Groups Complexity Cryptology {\bf 7},  117--133.
    Volume 7: Issue 2.


\bibitem{CHKPeik} David Cash, Dennis Hofheinz, Eike Kiltz, and Chris Peikert, {\it Bonsai Trees, or How to Delegate a Lattice Basis},   Advances in Cryptology – EUROCRYPT 2010, Springer Lecture Notes in Computer Science {\bf 6110}, pp 523--552.

\bibitem{CS} J.H. Conway and N.J.A. Sloane, {\it Sphere Packings, Lattices, and Groups}, 3rd ed., Grundlehren der mathematischen Wissenschafter {\bf 290}, Springer, New York (1999).

\bibitem{DRSm} W. Duke, Z. Rudnick, and P. Sarnak, {\it Density of integer points on affine homogeneous varieties}, Duke Math. Jour. {\bf 71} (1993), 143--179.

\bibitem{elkies}  Noam D. Elkies,  {\it A characterization of the $\Z^n$
lattice}, Math. Res. Lett. {\bf 2} (1995), 321–326.



\bibitem{GeisslerSmart} Katharina Gei{\ss}ler and Nigel P. Smart, {\it Computing the $M=UU^t$ integer matrix decomposition}, Cryptography and Coding 2003, Lect. Notes in Comp. Sci. {\bf 2898}, Springer, Berlin Heidelberg, 2003, 223--233.

\bibitem{GentrySz} C. Gentry and M. Szydlo, {\it Cryptanalysis  of  the  revised  NTRU  signature  scheme}, Advances in Cryptology—EUROCRYPT 2002, Lect. Notes in Comp. Sci. {\bf 2332}, Springer, Berlin, 2002, 299--320.
\url{http://www.szydlo.com/ntru-revised-full02.pdf}

\bibitem{Gerstein} Larry Gerstein, {\it Basic Quadratic Forms}, Graduate Studies in Mathematics {\bf 90}, Amer. Math. Soc.., Providence, RI, 2008.

\bibitem{GN} A. Gorodnik and A. Nevo, {\it The ergodic theory of lattice subgroups}, Annals of Mathematics Studies {\bf 172}, Princeton University Press, 2010.

\bibitem{recentlyshown} Christoph Hunkenschr\"oder, {\it Deciding whether a Lattice has an Orthonormal Basis is in co-NP}, arxiv:1910.03838


\bibitem{darkbases} Seungki Kim and Akshay Venkatesh, {\it The Behavior of Random Reduced Bases}, Int. Math. Res. Notices {\bf 2018}, pp. 6442--6480.

\bibitem{LLL} Arjen K. Lenstra, Jr., Hendrik W. Lenstra, and Laszlo Lovasz, {\it Factoring
polynomials with rational
coefficients},
Mathematische Annalen, {\bf 261}, pp. 513--534, (1982).


\bibitem{LenstraSilverberg0}  H.~W.~Lenstra Jr. and A. Silverberg, {\it Revisiting the Gentry-Szydlo Algorithm}, CRYPTO 2014,  Lecture Notes in Computer Science, {\bf 8616}, Springer, Berlin, pp.~280--296.


\bibitem{LenstraSilverberg}  H.~W.~Lenstra Jr. and A. Silverberg, {\it Lattices with symmetry}, Journal of Cryptology {\bf 30} (2017), 760-804.

\bibitem{LenstraSilverberg2}  H.~W.~Lenstra Jr. and A. Silverberg, {\it  Testing isomorphism of lattices over CM-orders}, SIAM Journal on Computing {\bf 48}, no. 4 (2019), 1300--1334.

\bibitem{MiccPeik} Daniele Micciancio and Chris Peikert, {\it Trapdoors for Lattices: Simpler, Tighter, Faster, Smaller}, Advances in Cryptology – EUROCRYPT 2012. Springer Lecture Notes in Computer Science {\bf 7237}, pp. 700--718.


\bibitem{NS} Phong Q. Nguyen and Damien Stehl\'e, {\it An LLL algorithm with quadratic complexity}, SIAM J. Comput, {\bf 39}, pp. 874--903 (2009).

\bibitem{DRSc} Thomas Plantard, Arnaud Sipasseuth, C\'edric Dumondelle, Willy Susilo, {\it DRS:~Diagonal dominant Reduction for lattice-based Signature}, NIST Post-Quantum Digital Signature Competition entry, \url{https://csrc.nist.gov/Projects/post-quantum-cryptography/Round-1-Submissions}

\bibitem{rivin} Igor Rivin, {\it How to pick a random integer matrix? (and other questions)}, Math. Comp. {\bf 85} (2016), 783--797.

\bibitem{BKZ} C.P. Schnorr, {\it A hierarchy of polynomial time lattice basis reduction algorithms}, Theoretical Computer Science {\bf 53} (1987), 201--224.

\bibitem{NISTRU} William Whyte and Lee Wilson, {\it Quantum Safety In
Certified Cryptographic
Modules}, \url{https://icmconference.org/wp-content/uploads/A21c-Whyte.pdf}

\end{thebibliography}
\end{document}